\documentclass[aps,prd,showpacs,notitlepage,nofootinbib,preprintnumbers,amsmath,amssymb]{revtex4-1}

\usepackage{graphics,graphicx}
\usepackage{dcolumn}
\usepackage{bm}
\usepackage{epsfig}
\usepackage[usenames]{color}
\usepackage{hyperref} 
\usepackage{mathbbol}
\usepackage{epstopdf}
\usepackage{simplewick}
\usepackage[utf8]{inputenc} 
\usepackage{slashed}
\usepackage{stackrel}
\usepackage{mathrsfs}
\usepackage{xcolor}
\usepackage{cancel}
\usepackage{mathtools}
\usepackage{multirow}
\usepackage{amsmath,amssymb,amsfonts,mathrsfs,bbold}
\usepackage[most]{tcolorbox}

\usepackage{physics}
\usepackage{subcaption}

\usepackage{ragged2e}
\DeclareCaptionJustification{justified}{\justifying}

\usepackage[normalem]{ulem}

\def\eq#1{{Eq.~(\ref{#1})}}
\def\fig#1{{Fig.~\ref{#1}}}
\def\sec#1{{Sec.~\ref{#1}}}
\def\tab#1{{Tab.~\ref{#1}}}
\newcommand{\ben}{\begin{eqnarray*}}
\newcommand{\een}{\end{eqnarray*}}
\newcommand{\un}[1]{\underline{#1}}

\newcommand{\pd}{\partial}
\newcommand{\bpsi}{\bar{\psi}}
\newcommand{\tord}{\textrm{T}}
\newcommand{\atord}{\bar{\textrm{T}}}

\newcommand{\llangle}{\Big\langle \!\! \Big\langle}
\newcommand{\rrangle}{\Big\rangle \!\! \Big\rangle}

\newcommand{\as}{\alpha_s}

\makeatletter
\DeclareRobustCommand{\cev}[1]{%
  {\mathpalette\do@cev{#1}}%
}
\newcommand{\do@cev}[2]{%
  \vbox{\offinterlineskip
    \sbox\z@{$\m@th#1 x$}%
    \ialign{##\cr
      \hidewidth\reflectbox{$\m@th#1\vec{}\mkern4mu$}\hidewidth\cr
      \noalign{\kern-\ht\z@}
      $\m@th#1#2$\cr
    }%
  }%
}
\makeatother

\begin{document}

\title{Spin-Spin Coupling at Small $x$: Worm-Gear and Pretzelosity TMDs}

\author{M. Gabriel Santiago}
  \email[Email: ]{gsantiago@sura.org}
	\affiliation{Center for Nuclear Femtography, SURA,
           1201 New York Ave. NW, Washington, DC 20005 USA}

\begin{abstract}
We study the small-$x$ asymptotics of the leading-twist quark transverse momentum dependent parton distribution functions (TMDs) which encode couplings between the polarization of the quarks and that of their parent hadron, with at least one of the two polarizations in the transverse direction: the two worm-gear TMDs $g_{1T}$ and $h_{1L}^{\perp}$, and the pretzelosity $h_{1T}^{\perp}$. We apply the Light Cone Operator Treatment originally developed in \cite{Kovchegov:2015pbl,Kovchegov:2017lsr, Kovchegov:2018znm, Kovchegov:2018zeq, Kovchegov:2021iyc}, finding that in the flavor non-singlet sector all three TMDs  reduce in the small-$x$ limit to previously known polarized dipole amplitudes, and thus the large-$N_c$, linearized, Double Logarithmic Approximation (DLA) asymptotics have all been solved for previously. For the worm-gear TMD $g_{1T}$ we find 
\begin{align}
    g_{1T}^{\textrm{NS}} (x \ll 1, k_T^2) \sim \left( \frac{1}{x} \right)^0 ,
\end{align}
while for the worm-gear TMD $h_{1L}^{\perp}$ we find
\begin{align}
    h_{1L}^{\perp \textrm{NS}} (x \ll 1, k_T^2) \sim \left( \frac{1}{x} \right)^{-1}.
\end{align}
Finally, for the pretzelosity TMD $h_{1T}^{\perp}$ we find
\begin{align} 
    h_{1 T}^{\perp \textrm{NS}} (x \ll 1, k_T^2) \sim \left( \frac{1}{x} \right)^{-1 + 2 \sqrt{\frac{\as N_c}{2\pi}}}.
\end{align}
We have compiled these asymptotics together with those of the other five flavor non-singlet, leading-twist quark TMDs into a single unified table.
\end{abstract}

\maketitle


\section{Introduction}
\label{sec_int}
The intersection of quantum chromodynamics (QCD) at small Bjorken $x$ and spin physics has been the topic of a surge of recent research activity \cite{Dominguez:2011wm, Dominguez:2011br, Kovchegov:2012ga,Kovchegov:2013cva,Zhou:2013gsa, Altinoluk:2014oxa, Boer:2015pni,Kovchegov:2015zha,Kovchegov:2015pbl, Dumitru:2015gaa, Szymanowski:2016mbq, Hatta:2016aoc, Hatta2016a,Hatta:2016khv,Boer:2016bfj,Balitsky:2016dgz, Kovchegov:2016zex, Kovchegov:2016weo, Kovchegov:2017jxc, Kovchegov:2017lsr, Dong:2018wsp, Benic:2018amn, Kovchegov:2018zeq, Kovchegov:2018znm, Chirilli:2018kkw, Altinoluk:2019wyu,Kovchegov:2019rrz, Boussarie:2019icw, Boussarie:2019vmk, Cougoulic:2019aja, Kovchegov:2020hgb, Cougoulic:2020tbc, Altinoluk:2020oyd, Kovchegov:2020kxg, Kovchegov:2021iyc, Chirilli:2021lif, Altinoluk:2021lvu, Adamiak:2021ppq, Kovchegov:2021lvz,Bondarenko:2021rbp,Banu:2021cla, Cougoulic:2022gbk,Kovchegov:2022kyy,Benic:2022qzv,Hatta:2022bxn,Borden:2023ugd, Li:2023tlw}. In particular, the inclusion of sub-eikonal and sub-sub-eikonal corrections \cite{Altinoluk:2014oxa,Kovchegov:2015pbl,Balitsky:2016dgz, Hatta:2016aoc, Kovchegov:2016zex, Kovchegov:2016weo, Kovchegov:2017jxc, Kovchegov:2017lsr, Kovchegov:2018znm, Kovchegov:2018zeq,Chirilli:2018kkw, Altinoluk:2019wyu, Kovchegov:2019rrz, Cougoulic:2019aja, Kovchegov:2020hgb, Cougoulic:2020tbc, Altinoluk:2020oyd, Chirilli:2021lif, Adamiak:2021ppq, Altinoluk:2021lvu, Kovchegov:2021lvz, Kovchegov:2021iyc, Cougoulic:2022gbk, Li:2023tlw}  to the saturation/Color Glass Condensate (CGC) framework \cite{Gribov:1984tu,Iancu:2003xm,Weigert:2005us,JalilianMarian:2005jf,Gelis:2010nm,Albacete:2014fwa,Kovchegov:2012mbw,Morreale:2021pnn} has been explored in detail, both in the study of transverse momentum dependent parton distribution functions (TMDs) at small-$x$ as well as in the direct study of quark and gluon propagators through the strong gluon field background of the CGC. This extension of the saturation framework allows one to go beyond the Collins-Soper-Sterman evolution equations \cite{Collins:1981uw,Collins:1981uk,Collins:1981va,Collins:1984kg,Collins:1989gx} which resum logarithms of the hard scale $Q^2$, and instead resum logarithms of $1/x$ to predict the small-$x$ behavior of TMDs. Using the Light Cone Operator Treatment (LCOT) originally developed in \cite{Kovchegov:2015pbl,Kovchegov:2017lsr, Kovchegov:2018znm, Kovchegov:2018zeq, Kovchegov:2021iyc}, the evolution of several leading-twist quark and gluon TMDs has been studied by constructing polarized dipole amplitudes containing spin-dependent sub-eikonal and sub-sub-eikonal corrections to the usual eikonal, small-$x$ dipole amplitudes. Including the unpolarized quark TMD, which has small-$x$ behavior determined by the evolution equation for the Reggeon \cite{Kirschner:1983di,Kirschner:1985cb,Kirschner:1994vc,Kirschner:1994rq,Griffiths:1999dj,Itakura:2003jp}, five of the eight leading-twist quark TMDs have known small-$x$ asymptotics. In this work, we will study the small-$x$ asymptotics of the remaining three leading-twist quark TMDs, the two worm-gear TMDs, $g_{1T}$ and $h_{1L}^{\perp}$, and the pretzelosity $h_{1T}^{\perp}$. These three TMDs vanish upon integration over the quark transverse momentum $k_T$, and encode various couplings between the polarization of the quarks with the polarization of their parent hadron \cite{Mulders:1995dh}. Studying their small-$x$ asymptotics will give us a window into this spin-spin coupling structure in the gluon dominated, high energy regime of QCD. We will apply the LCOT to all three TMD operator definitions, then find the asymptotics of the flavor non-singlet TMDs in the large-$N_c$ limit, further taking the linearized double logarithmic approximation (DLA).

\begin{table}[h] 
    \def\arraystretch{1.25}
    \centering
    \begin{tabular}{| c | c | c | c | c |} 
    \hline
    \multicolumn{5}{|c|}{Leading Twist Quark TMDs} \\ \hline
      & & \multicolumn{3}{|c|}{Quark Polarization} \\ \cline{3-5}
      & & U & L & T \\ \hline
      \multirow{3}{2.0cm}{Nucleon Polarization} & U & $f_1^{\textrm{NS}} \sim x^{-\sqrt{2 \as C_F / \pi}}$ &  & $h_1^{\perp \textrm{NS}} \sim x$ \\ \cline{2-5}
      & L & & $g_1^{\textrm{NS}} \sim x^{-\sqrt{\as N_c / \pi}}$ & $h_{1L}^{\perp \textrm{NS}} \sim x$ \\ \cline{2-5}
      & T & $f_{1T}^{\perp \textrm{NS}} \sim C_{\mathcal{O}} x^{-1} + C_1 x^{-3.4 \sqrt{\as N_c / 4 \pi}}$ & $g_{1T}^{\textrm{NS}} \sim x^{0}$ & $h_1^{\textrm{NS}} \sim h_{1T}^{\perp \textrm{NS}} \sim x^{1 - 2 \sqrt{\as N_c / 2 \pi}}$ \\ \cline{2-5}
      \hline
    \end{tabular} 
    \caption{The collected leading small-$x$ asymptotics for the leading-twist flavor non-singlet quark TMDs. The intercept of the unpolarized quark TMD $f_1$ was found from the evolution equation for the Reggeon \cite{Kirschner:1983di,Kirschner:1985cb,Kirschner:1994vc,Kirschner:1994rq,Griffiths:1999dj,Itakura:2003jp} and also with the Infrared Evolution Equation (IREE) \cite{Ermolaev:1995fx}, while the remaining seven intercepts have been calculated using the LCOT in \cite{Kovchegov:2015pbl,Kovchegov:2017lsr, Kovchegov:2018znm, Kovchegov:2018zeq, Kovchegov:2022kyy} and this work. The intercept for the flavor non-singlet helicity TMD was also found in \cite{Bartels:1995iu} using the IREE and agrees with the result found using the LCOT. Finally, the intercept of the flavor non-singlet transversity TMD found in the LCOT matches that found in \cite{Kirschner:1996jj}.}
    \label{tab_tmds}
\end{table}

The structure of the remainder of this paper is as follows: we will start in \sec{sec_g1t} with the worm-gear $g_{1T}$ function, finding that in the small-$x$ limit it reduces to the same polarized dipole amplitudes which are present in the sub-eikonal contribution to the Sivers function \cite{Kovchegov:2022kyy}, but with a crucial change from the real part of a correlator to the imaginary part. This alters the initial conditions for the evolution of the polarized dipole amplitudes, and leads to the same evolution equations as for the sub-sub-eikonal Boer-Mulders function \cite{Kovchegov:2022kyy}, yielding $x$-independent, exactly sub-eikonal asymptotics 
\begin{align}
    g_{1T}^{\textrm{NS}} (x \ll 1, k_T^2) \sim \left( \frac{1}{x} \right)^0 .
\end{align}
Next in \sec{sec_h1l} we will study the worm-gear $h_{1L}^{\perp}$ function. It will turn out to share analogous polarized dipole amplitudes to the sub-sub-eikonal Boer-Mulders function \cite{Kovchegov:2022kyy}, leading to exactly sub-sub-eikonal scaling
\begin{align}
    h_{1L}^{\perp \textrm{NS}} (x \ll 1, k_T^2) \sim \left( \frac{1}{x} \right)^{-1},
\end{align}
giving with $g_{1T}$ two examples of time reversal even (T-even) TMDs which receive essentially no corrections to their naive scaling with $x$ in the linearized evolution regime. We continue on to \sec{sec_pretz} where we study the pretzelosity TMD, finding that it shares polarized dipole amplitudes with the transversity TMD \cite{Kovchegov:2018zeq} and has the same asymptotic scaling as
\begin{align} 
    h_{1 T}^{\perp \textrm{NS}} (x \ll 1, k_T^2) \sim \left( \frac{1}{x} \right)^{-1 + 2 \sqrt{\frac{\as N_c}{2\pi}}},
\end{align}
coming from the evolution equation for the Reggeon \cite{Kirschner:1983di,Kirschner:1985cb,Kirschner:1994vc,Kirschner:1994rq,Griffiths:1999dj,Itakura:2003jp}.
Finally, in \sec{sec_conc} we summarize our results and point out some directions for future work. We have collected these asymptotics together with those for the other leading-twist quark TMDs in \tab{tab_tmds}. 

Throughout this paper we will make use of light-cone coordinates $u = (u^+ = x^0 + x^3, u^- = u^0 - u^3, \un{u})$, labelling the transverse part of a four-vector $u$ as $\un{u}$ except in the case of an integral measure, where it will be denoted as $u_{\perp}$, and in the case of the quark transverse momentum argument of a TMD, where we will use the conventional label $k_T$. We will also make use of Brodsky-Lepage (BL) spinors \cite{Lepage:1980fj}, specifically in the plus-minus reversed form defined as \cite{Kovchegov:2018znm,Kovchegov:2018zeq}
\begin{align}\label{anti BLspinors}
u_\sigma (p) = \frac{1}{\sqrt{p^-}} \, [p^- + m \, \gamma^0 +  \gamma^0 \, {\un \gamma} \cdot {\un p} ] \,  \rho (\sigma), \ \ \ v_\sigma (p) = \frac{1}{\sqrt{p^-}} \, [p^- - m \, \gamma^0 +  \gamma^0 \, {\un \gamma} \cdot {\un p} ] \,  \rho (-\sigma),
\end{align}
with $p^\mu = \left( \frac{{\un p}^2+ m^2}{p^-}, p^-, {\un p} \right)$ and
\begin{align}
  \rho (+1) \, = \, \frac{1}{\sqrt{2}} \, \left(
  \begin{array}{c}
      1 \\ 0 \\ -1 \\ 0
  \end{array}
\right), \ \ \ \rho (-1) \, = \, \frac{1}{\sqrt{2}} \, \left(
  \begin{array}{c}
        0 \\ 1 \\ 0 \\ 1
  \end{array}
\right) .
\end{align}
We will refer to these as anti BL spinors.

\section{Worm-Gear $g_{1T}$}
\label{sec_g1t}

We begin with the worm-gear TMD $g_{1T}$, which can be interpreted as the number density of longitudinally polarized quarks within a transversely polarized proton. It is defined as \cite{Meissner:2007rx}
\begin{align}
\frac{k_T \vdot S_P}{M_P} g_{1T} (x, k_T^2) = \int \frac{\dd{r}^- \dd[2]{r}_{\perp}}{2(2\pi)^3} \bra{P, S_P} \bpsi (0) \mathcal{U} [0, r] \frac{\gamma^+ \gamma_5}{2} \psi(r)  \ket{P, S_P} ,
\end{align}
where the proton spin $S_P$ is in the transverse plane and we take the future pointing semi-inclusive deep inelastic scattering (SIDIS) staple gauge link $U [0,r] = V_{\un{0}} [0, \infty] V_{\un{r}} [\infty, r^-]$ with fundamental light-cone Wilson lines defined as
\begin{align}\label{Vline}
V_{\un{x}} [x^-_f,x^-_i] = \mathcal{P} \exp \left[ \frac{ig}{2} \int\limits_{x^-_i}^{x^-_f} \dd{x}^- A^+ (0^+, x^-, \un{x}) \right] .
\end{align} We have dropped the transverse pieces of the gauge link out at infinity, so we must work in a non-singular gauge. Here we will take the proton `target' to be moving along the $x^+$ direction, and will work in $A^- = 0$ gauge for our calculations. We apply the LCOT \cite{Kovchegov:2018znm,Kovchegov:2018zeq}, rewriting the matrix element as a small-$x$ quasi-classical average in the `target' proton state \cite{McLerran:1993ni,McLerran:1993ka,McLerran:1994vd,Kovchegov:1996ty,Balitsky:1997mk,Balitsky:1998ya,Kovchegov:2019rrz}
\begin{align}
    \frac{k_T \vdot S_P}{M_P} g_{1T} (x, k_T^2) \subset \frac{2 p_1^+}{2 (2 \pi)^3} \sum_X \int \dd{\xi}^- \dd[2]{\xi}_{\perp} \dd{\zeta}^- \dd[2]{\zeta}_{\perp} e^{i k \vdot (\xi - \zeta)} \left[ \frac{\gamma^+ \gamma_5}{2} \right]_{\alpha \beta} \langle \bpsi_{\alpha} (\xi) V_{\un{\xi}} [\xi^-, \infty] \ket{X} \bra{X} V_{\un{\zeta}} [\infty, \zeta^-] \psi_{\beta} (\zeta) \rangle .
\end{align}
Here we have inserted a sum over the complete set of states $\ket{X}$, and the outer angular brackets indicate the quasi-classical averaging over the proton wave function.

\begin{figure}[h]
\centering
\includegraphics[width=0.5\linewidth]{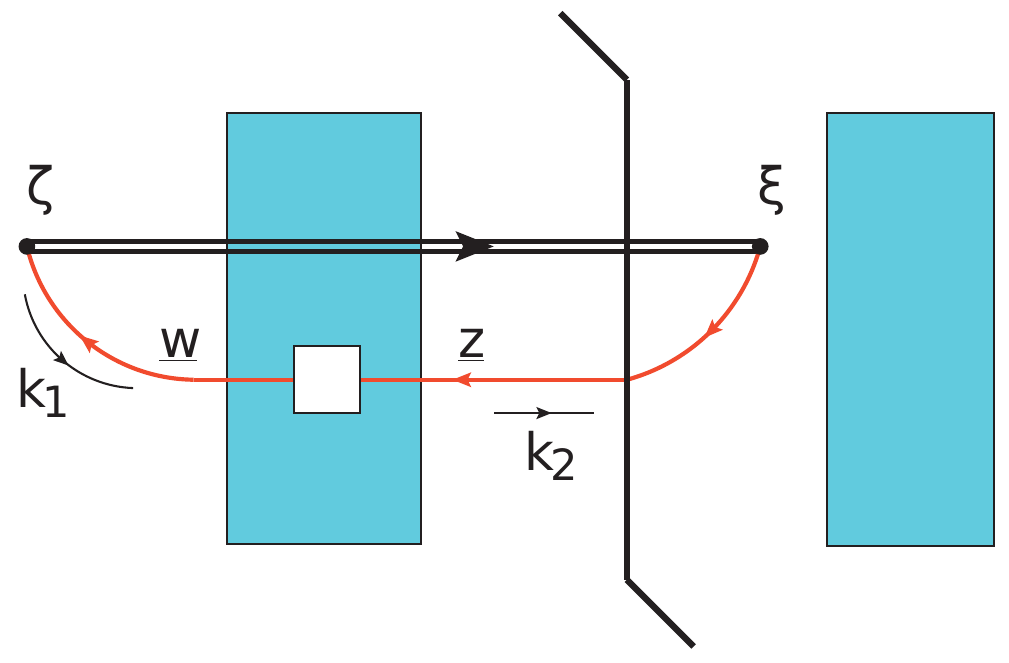}  
\caption{Example of the class of diagrams which give the leading sub-eikonal and sub-sub-eikonal corrections for the TMDs considered here. The anti-quark propagates from the position $\zeta$ to $\un{w}$ with momentum $k_1$, undergoes a sub-eikonal interaction with the proton which changes its transverse position from $\un{w}$ on the left of the shock wave to $\un{z}$ on the right of the shock wave. The anti-quark then propagates from $\un{z}$ to the position $\xi$ with momentum $k_2$. The shock wave is denoted by the blue (grey) rectangle, while the sub-(sub-) eikonal interaction with the shock wave is denoted by the white box. The double line represents the eikonal Wilson line encoding the interactions of the quark in the dipole.}
\label{FIG:diagbdet}
\end{figure}

One can show that the eikonal contribution to this TMD vanishes, just as for the other quark spin dependent leading-twist quark TMDs. The leading contribution then comes from diagrams of the class shown in \fig{FIG:diagbdet}, where we have taken $\ket{X}$ to be an anti-quark state and added a sub-eikonal spin-dependent exchange (the white box) onto the Wilson line which encodes the interactions between the anti-quark and the target as it passes through the shock wave (the blue (grey) rectangle). Evaluating these diagrams (cf. \cite{Kovchegov:2018znm,Kovchegov:2018zeq}) yields the sub-eikonal contribution to the worm-gear function as
\begin{align}\label{seg1t_1}
    \frac{k_T \vdot S_P}{M_P} g_{1T} (x, k_T^2) |_{\textrm{sub-eik.}} &\subset \frac{-2 p_1^+}{2(2 \pi)^3} \int \dd[2]{\zeta}_{\perp} \dd[2]{w}_{\perp} \dd[2]{z}_{\perp} \frac{\dd{k}_1^- \dd[2]{k}_{1 \perp}}{(2 \pi)^3}  \frac{e^{i \un{k}_1  \vdot (\un{w} - \un{\zeta}) + i \un{k} \vdot (\un{z} - \un{\zeta})} \theta(k_1^-)}{(x p_1^+ k_1^- + \un{k}_1^2) (x p_1^+ k_1^- + \un{k}^2)}  \\ 
    &\times \sum_{\sigma_1, \sigma_2} \bar{v}_{\sigma_2} (k_2) \frac{\gamma^+ \gamma_5}{2} v_{\sigma_1} (k_1) \langle \tord V_{\un{\zeta}}^{ij} \bar{v}_{\sigma_1} (k_1) \hat{V}_{\un{z},\un{w}}^{\dagger \ ji} v_{\sigma_2} (k_2) \rangle \Big|_{k_1^- = k_2^-, k_1^2 = k_2^2 = 0, \un{k}_2 = -\un{k}} + c.c. .  \notag
\end{align}
We now need the anti BL spinor product
\begin{align} \label{antivprod}
    \bar{v}_{\sigma_2} (k_2) \frac{\gamma^+ \gamma_5}{2} v_{\sigma_1} (k_1) &= \frac{1}{2} \frac{\sigma_2 \delta_{\sigma_2, \sigma_1} (\un{k}_2 \vdot \un{k}_1) - i \delta_{\sigma_2,  \sigma_1} (\un{k}_2 \cross \un{k}_1)}{\sqrt{k_2^- k_1^-}},
\end{align}
as well as the replacement \cite{Cougoulic:2022gbk,Kovchegov:2022kyy} 
\begin{align}\label{V_repl}
\bar{v}_{\chi_1} (k_1) \left( \hat{V}_{\un{z},{\un w}}^\dagger \right)^{ji}  v_{\chi_2} (k_2)  \to 2 \sqrt{k_1^- \, k_2^-}  \,  \int d^2 z_\perp \, \left( V^{\textrm{pol} \, \dagger}_{{\un z}, {\un w}; \chi_2 , \chi_1} \right)^{ji} ,
\end{align}
where we have the polarized Wilson line
\begin{align}\label{Vphase&mag}
V_{\un{x}, \underline{y}; \chi', \chi}^{\textrm{pol}} = \delta_{\chi, \chi'} \, V_{\un{x}, \un{y}}^{\textrm{phase}} + \delta_{\chi, - \chi'} \, V_{\un{x}, \un{y}}^{\textrm{mag}}
\end{align}
with
\begin{subequations}\label{V_subeik_1}
\begin{align}
& V_{\un{x}, \un{y}}^{\textrm{phase}} = - \frac{i \, p_1^+}{2 \, s}  \int\limits_{-\infty}^{\infty} d{z}^- d^2 z \ V_{\un{x}} [ \infty, z^-] \, \delta^2 (\un{x} - \un{z}) \, \cev{D}_z^i  \vec{D}^i_z \, V_{\un{y}} [ z^-, -\infty] \, \delta^2 (\un{y} - \un{z}) \label{V_subeik_1a} \\ 
& - \frac{g^2 \, p_1^+}{4 \, s} \, \delta^2 (\un{x} - \un{y})  \, \int\limits_{-\infty}^{\infty} d{z}_1^- \int\limits_{z_1^-}^\infty d z_2^-  \ V_{\un{x}} [ \infty, z_2^-] \,  t^b \, \psi_{\beta} (z_2^-,\un{x}) \, U_{\un{x}}^{ba} [z_2^-,z_1^-] \, \left[ \frac{\gamma^+}{2}  \right]_{\alpha \beta} \bar{\psi}_\alpha (z_1^-,\un{x}) \, t^a \, V_{\un{x}} [ z_1^-, -\infty] , \notag \\
& V_{\un{x}, \underline{y}}^{\textrm{mag}} = \frac{i \, g \, p_1^+}{2 \, s} \delta^2 (\un{x} - \un{y}) \,  \int\limits_{-\infty}^{\infty} d{z}^- \ V_{\un{x}} [ \infty, z^-] \, F^{12} \, V_{\un{x}} [ z^-, -\infty] \label{V_subeik_1b} \\ 
& - \frac{g^2 \, p_1^+}{4 \, s} \, \delta^2 (\un{x} - \un{y})  \, \int\limits_{-\infty}^{\infty} d{z}_1^- \int\limits_{z_1^-}^\infty d z_2^-  \ V_{\un{x}} [ \infty, z_2^-] \,  t^b \, \psi_{\beta} (z_2^-,\un{x}) \, U_{\un{x}}^{ba} [z_2^-,z_1^-] \, \left[ \frac{\gamma^+ \gamma^5}{2} \right]_{\alpha \beta} \bar{\psi}_\alpha (z_1^-,\un{x}) \, t^a \, V_{\un{x}} [ z_1^-, -\infty] . \notag 
\end{align}
\end{subequations}
Here $\psi$ and $\bar{\psi}$ are the background quark and anti-quark fields of the target hadron, $F^{12}$ is a component of the background gluon field strength tensor, and $\vec{D}^i = \vec{\pd}_i - i g A^i, \cev{D}^i = \cev{\pd}_i + i g A^i$ are the left and right acting fundamental representation covariant derivatives. We have also introduced the adjoint representation Wilson line \begin{align}\label{Uline}
U_{\un{x}} [x^-_f,x^-_i] = \mathcal{P} \exp \left[ \frac{ig}{2} \int\limits_{x^-_i}^{x^-_f} \dd{x}^- {\cal A}^+ (0^+, x^-, \un{x}) \right]
\end{align}
with $U_{\un{x}} \equiv U_{\un{x}} [\infty,-\infty]$. Plugging \eq{antivprod} and \eq{V_repl} into \eq{seg1t_1} yields 
\begin{align} \label{seg1t_2}
    \frac{k_T \vdot S_P}{M_P} g_{1T} (x, k_T^2) |_{\textrm{sub-eik.}} &\subset \frac{2 p_1^+}{(2 \pi)^3} \int \dd[2]{\zeta}_{\perp} \dd[2]{w}_{\perp} \dd[2]{z}_{\perp} \frac{\dd{k}_1^- \dd[2]{k}_{1 \perp}}{(2 \pi)^3} \theta(k_1^-) \frac{1}{(x p_1^+ k_1^- + \un{k}_1^2) (x p_1^+ k_1^- + \un{k}^2)} \\
    \times \Big{\langle} i \un{k}_1 \cross \un{k}&  \left( e^{i \un{k}_1 \vdot (\un{w} - \un{\zeta}) + i \un{k} \vdot ( \un{z} - \un{\zeta})} \tord \tr \left[ V_{\un{\zeta}} V_{\un{z},\un{w}}^{\textrm{phase} \dagger} \right] - e^{- i \un{k}_1 \vdot (\un{w} - \un{\zeta}) - i \un{k} \vdot ( \un{z} - \un{\zeta})} \atord \tr \left[ V_{\un{z},\un{w}}^{\textrm{phase}} V_{\un{\zeta}}^{\dagger} \right] \right) \notag \\
    &+ e^{i (\un{k}_1 + \un{k}) \vdot (\un{w} - \un{\zeta})} \un{k}_1 \vdot \un{k} \left( \tord \tr \left[ V_{\un{\zeta}} V_{\un{z},\un{w}}^{\textrm{mag} \dagger} \right] + \atord \tr \left[ V_{\un{z},\un{\zeta}}^{\textrm{mag}} V_{\un{w}}^{\dagger} \right] \right) \Big{\rangle}  , \notag
\end{align}
where we have interchanged $\un{w}$ and $\un{\zeta}$ in the last trace. We can recognize in \eq{seg1t_2} the exact same Wilson line correlators which are present in the sub-eikonal contribution to the Sivers function \cite{Kovchegov:2022kyy}. There the left hand side (LHS) contained a cross product $k_T \cross S_P$ and the right hand side (RHS) had $\un{k}_1 \vdot \un{k}$ multiplying the $V_{\un{z},\un{w}}^{\textrm{phase} \dagger}$ correlator and $i \un{k}_1 \cross \un{k}$ multiplying the $V_{\un{z},\un{w}}^{\textrm{mag} \dagger}$ correlator. The equation above changes the LHS cross product in the Sivers case to a dot product $k_T \vdot S_P$ here, and exchanges the $\un{k}_1, \un{k}$ structures on the RHS. Thus, as in the case of the Sivers function, the polarized dipole amplitude containing $V_{\un{z},\un{w}}^{\textrm{phase} \dagger}$ must bring in a factor of $\epsilon^{ij}$ and be linear in $S_P$ once integrated over impact parameter, while the polarized dipole amplitude containing $V_{\un{z},\un{w}}^{\textrm{mag} \dagger}$ must have no factor of $\epsilon^{ij}$ and be linear in $S_P$ once integrated over impact parameter. Having the same tensor structures in front of the impact parameter integrated polarized dipole amplitudes allows us to apply the results for the sub-eikonal contribution to the Sivers function to $g_{1T}$, taking the flavor non-singlet TMD, defined as the difference between the quark and anti-quark TMDs
\begin{align}
    g_{1T}^{\textrm{NS}} = g_{1T}^q - g_{1T}^{\bar{q}} ,
\end{align}
to obtain
\begin{align} \label{seg1t_fin}
    \frac{k_T \vdot S_P}{M_P} g_{1T}^{\textrm{NS}} (x, k_T^2) |_{\textrm{sub-eik.}} &= \frac{8 N_c p_1^+}{(2\pi)^3} \int \dd[2]{\zeta}_{\perp} \dd[2]{w}_{\perp} \frac{\dd[2]{k}_{1 \perp}}{(2 \pi)^3} e^{i ( \un{k} + \un{k}_1) \vdot ( \un{w} - \un{\zeta})}  \frac{1}{\un{k}_1^2 \un{k}^2} \\
    &\times \int\limits_{\Lambda/s}^1 \frac{\dd{z}}{z} \left[ i \un{k}_1 \cross \un{k} (\un{k} - \un{k}_1)^i G_{\un{w} \un{\zeta}}^{i} + i \un{k}_1 \cross \un{k} \, G_{\un{w} \un{\zeta}}^{[2]} + \un{k}_1 \vdot \un{k} \, G_{\un{w} \un{\zeta}}^{\textrm{mag}} \right], \notag
\end{align}
with
\begin{subequations}\label{seg1t_dips}
    \begin{align}
        &G_{\un{w} \un{\zeta}}^{i} = \frac{1}{2 N_c} \textrm{Im} \llangle \tord \tr \left[ V_{\un{\zeta}} V_{\un{w}}^{i \, \dagger} \right] - \tord \tr \left[ V_{\un{w}}^i V_{\un{\zeta}}^{\dagger} \right] \rrangle, \label{seg1t_dips_fin1} \\
        &G_{\un{w} \un{\zeta}}^{[2]} = \frac{1}{2 N_c} \textrm{Re} \llangle \tord \tr \left[ V_{\un{\zeta}} V_{\un{w}}^{[2] \, \dagger} \right] - \tord \tr \left[ V_{\un{w}}^{[2]} V_{\un{\zeta}}^{\dagger} \right] \rrangle, \label{seg1t_dips_fin[2]} \\
        &G_{\un{w} \un{\zeta}}^{\textrm{mag}} = \frac{1}{2 N_c} \textrm{Im} \llangle \tord \tr \left[ V_{\un{\zeta}} V_{\un{w}}^{\textrm{mag} \, \dagger} \right] - \tord \tr \left[ V_{\un{w}}^\textrm{mag} V_{\un{\zeta}}^{\dagger} \right] \rrangle \label{seg1t_dips_fin2}
    \end{align}
\end{subequations}
where the double angle brackets scale out a power of energy, $\llangle ... \rrangle = zs \Big{\langle} ... \Big{\rangle}$. Here we have introduced the polarized Wilson lines
\begin{subequations}
    \begin{align}
        & V_{\un{x}}^i = \frac{p_1^+}{4 \, s}  \int\limits_{-\infty}^{\infty} d{z}^- \ V_{\un{x}} [ \infty, z^-] \, \left[ \vec{D}^i_x -  \cev{D}_x^i \right]  \, V_{\un{x}} [ z^-, -\infty], \label{V_pol_i}\\
        & V_{\un{x}; \un{k}, \un{k}_1}^{[2]} = \frac{i \, p_1^+}{8 \, s}  \int\limits_{-\infty}^{\infty} d{z}^- \ V_{\un{x}} [ \infty, z^-] \, \left[ (\vec{D}^i_x -  \cev{D}_x^i)^2 - (k_1^i -  k^i)^2 \right]  \, V_{\un{x}} [ z^-, -\infty] \label{V_pol_2} \\
        & - \frac{g^2 \, p_1^+}{4 \, s} \, \int\limits_{-\infty}^{\infty} d{z}_1^- \int\limits_{z_1^-}^\infty d z_2^-  \ V_{\un{x}} [ \infty, z_2^-] \,  t^b \, \psi_{\beta} (z_2^-,\un{x}) \, U_{\un{x}}^{ba} [z_2^-,z_1^-] \, \left[ \frac{\gamma^+}{2}  \right]_{\alpha \beta} \bar{\psi}_\alpha (z_1^-,\un{x}) \, t^a \, V_{\un{x}} [ z_1^-, -\infty] \notag
    \end{align}
\end{subequations}
which were contained in \eq{V_subeik_1a}.

The only difference between the dipole amplitudes in \eq{seg1t_dips} and those in the sub-eikonal contribution to the flavor non-singlet Sivers function \cite{Kovchegov:2022kyy} is the Im operator acting on the correlators containing $V_{\un{x}}^i$ and $V_{\un{x}}^{\textrm{mag}}$ in \eq{seg1t_dips_fin1} and \eq{seg1t_dips_fin2}, rather than a Re operator as in the Sivers case, and vice versa for the correlator containing $V_{\un{x}}^{[2]}$ in \eq{seg1t_dips_fin[2]}. This change does not affect any of the evolution calculations at the level of the polarized Wilson line correlators, but it significantly changes the initial conditions for the polarized dipole amplitudes. We will argue below that the only polarized dipole amplitude which contributes to the linearized, DLA small-$x$ asymptotics of the worm-gear $g_{1T}$ is $G_{\un{w} \un{\zeta}}^{[2]}$, with all other amplitudes dropping out due to having initial conditions equal to zero and having no mixing with $G_{\un{w} \un{\zeta}}^{[2]}$ under evolution. This means we can simplify the worm-gear TMD further as
\begin{align} \label{seg1t_fin_simp}
    \frac{k_T \vdot S_P}{M_P} g_{1T}^{\textrm{NS}} (x, k_T^2) |_{\textrm{sub-eik.}} &= \frac{8 N_c p_1^+}{(2\pi)^3} \int \dd[2]{\zeta}_{\perp} \dd[2]{w}_{\perp} \frac{\dd[2]{k}_{1 \perp}}{(2 \pi)^3} e^{i ( \un{k} + \un{k}_1) \vdot ( \un{w} - \un{\zeta})}  \frac{i \un{k}_1 \cross \un{k}}{\un{k}_1^2 \un{k}^2}  \int\limits_{\Lambda/s}^1 \frac{\dd{z}}{z}  \, G_{\un{w} \un{\zeta}}^{[2]} \notag \\
    = \frac{8 N_c p_1^+}{(2\pi)^3} &\int \dd[2]{x}_{10} \frac{\dd[2]{k}_{1 \perp}}{(2 \pi)^3} e^{i ( \un{k} + \un{k}_1) \vdot \un{x}_{10}}  \frac{i \un{k}_1 \cross \un{k}}{\un{k}_1^2 \un{k}^2}  \int\limits_{\Lambda/s}^1 \frac{\dd{z}}{z} (\un{x}_{10} \cross \un{S}_{P}) \, G^{[2]} (x_{10}^2, z),
\end{align}
where we have introduced $\un{x}_{10} = \un{w} - \un{\zeta}$ and integrated over impact parameter, using the $k_T \vdot S_P$ term on the LHS to deduce the factor of $\un{x}_{10} \cross \un{S}_P$ on the RHS in the last line.

In order to argue that this simplification is valid, it will be helpful to separately consider the quark exchange and gluon exchange terms within the polarized dipole amplitudes. For $G_{\un{w} \un{\zeta}}^i$ this is trivial, as the only sub-eikonal operator entering this dipole amplitude through the Wilson line $V_{\un{x}}^i$ is a gluon exchange operator. For $G_{\un{w} \un{\zeta}}^{[2]}$, we can write
\begin{align}
    G_{\un{w} \un{\zeta}}^{[2]} = G_{\un{w} \un{\zeta}}^{[2] \, g} + G_{\un{w} \un{\zeta}}^{[2] \, q},
\end{align}
with $G_{\un{w} \un{\zeta}}^{[2] \, g}$ containing the operators in the first line of \eq{V_pol_2} and $G_{\un{w} \un{\zeta}}^{[2] \, q}$ containing those in the second line. For $G_{\un{w} \un{\zeta}}^{\textrm{mag}}$ we can make a similar decomposition as
\begin{align}
    G_{\un{w} \un{\zeta}}^{\textrm{mag}} = G_{\un{w} \un{\zeta}}^{\textrm{mag} \, g} + G_{\un{w} \un{\zeta}}^{\textrm{mag} \, q},
\end{align}
with $G_{\un{w} \un{\zeta}}^{\textrm{mag} \, g}$ containing the operators in the first line of \eq{V_subeik_1b}, and $G_{\un{w} \un{\zeta}}^{\textrm{mag} \, q}$ containing those in the second line. Now we consider the gluon exchange component of the operator for the three polarized dipole amplitudes defined in \eq{seg1t_dips}. From \cite{Kovchegov:2021iyc,Kovchegov:2022kyy} we know that, for a transversely polarized target, the gluon exchange dipole amplitudes $G_{\un{w} \un{\zeta}}^{[2] \, g}$ and $G_{\un{w} \un{\zeta}}^{\textrm{mag} \, g}$ have zero initial conditions. For $G_{\un{w} \un{\zeta}}^i$, we consider the correlator 
\begin{align} \label{odd_i_corr}
    \left< \tord \tr \left[ V_{\un{\zeta}} V_{\un{w}}^{i \, \dagger} \right] - \tord \tr \left[  V_{\un{w}}^{i} V_{\un{\zeta}}^{\dagger} \right] \right> .
\end{align}
From \cite{Kovchegov:2021iyc,Kovchegov:2022kyy} we know that for a transversely polarized target this correlator has a nonzero real term in its initial value. This real contribution comes from a triple gluon exchange between the dipole and the target, taking the gluons to be in the symmetric $d^{abc} = 2 \tr \left[ \{ t^a, t^b\} t^c \right]$ color representation, similar to how the odderon (at leading order) comes from an eikonal triple gluon exchange in the color symmetric representation \cite{Kovchegov:2003dm,Hatta:2005as,Kovner:2005qj,Jeon:2005cf,Kovchegov:2012ga}. While the odderon gives the imaginary part of the eikonal dipole amplitude (proportional to the eikonal Wilson line correlator $\langle \tr \left[ V_{\un{x}} V_{\un{y}}^{\dagger} \right] \rangle$), the similar initial condition for \eq{odd_i_corr} is real due to an extra factor of $i$ entering the sub-eikonal operator in \eq{V_pol_i}. In order to avoid a cancellation between a gluon exchange amplitude and the same amplitude with quark and anti-quark interchanged as in \eq{odd_i_corr}, the correlator must be odderon-like, and the extra $i$ will make this a pure real contribution. Thus we find that $G_{\un{w} \un{\zeta}}^i$ has an initial condition of zero due to the Im operator in \eq{seg1t_dips_fin1}.

As all of the initial conditions for polarized gluon exchange dipole amplitudes are zero, and we are studying the flavor non-singlet worm-gear TMD, resumming polarized gluon emissions will not contribute to evolution because no flavor information would be communicated between the dipole and the target. One can consider a combination of polarized gluon and polarized quark emissions, but these terms would be suppressed by factors of $N_c$, so we will not include polarized gluon emissions in the linearized, large-$N_c$, DLA evolution equations which we consider here (cf \cite{Kovchegov:2016zex,Kovchegov:2018zeq}). As we will see, the quark exchange dipole amplitudes do not contribute to $G_{\un{w} \un{\zeta}}^i$, $G_{\un{w} \un{\zeta}}^{[2] \, g}$ and $G_{\un{w} \un{\zeta}}^{\textrm{mag} \, g}$ through evolution, so these dipole amplitudes fully drop out of $g_{1T}$.

Now we turn to the quark exchange terms $G_{\un{w} \un{\zeta}}^{\textrm{mag} \, q}$ and $G_{\un{w} \un{\zeta}}^{[2] \, q}$. One can show that the Wilson line correlators in each of these polarized dipole amplitudes has pure real initial conditions (see Sec. III of \cite{Kovchegov:2016zex} or App. B of \cite{Kovchegov:2018znm}), so $G_{\un{w} \un{\zeta}}^{\textrm{mag} \, q}$ will have an initial condition of zero while $G_{\un{w} \un{\zeta}}^{[2] \, q}$ can be nonzero. We will find that these two dipole amplitudes do not mix under evolution, so $G_{\un{w} \un{\zeta}}^{\textrm{mag} \, q}$ drops out of the TMD and we are indeed left with only $G_{\un{w} \un{\zeta}}^{[2] \, q}$ entering in \eq{seg1t_fin_simp}.

\begin{figure}[ht]
\centering
\includegraphics[width= 0.8 \linewidth]{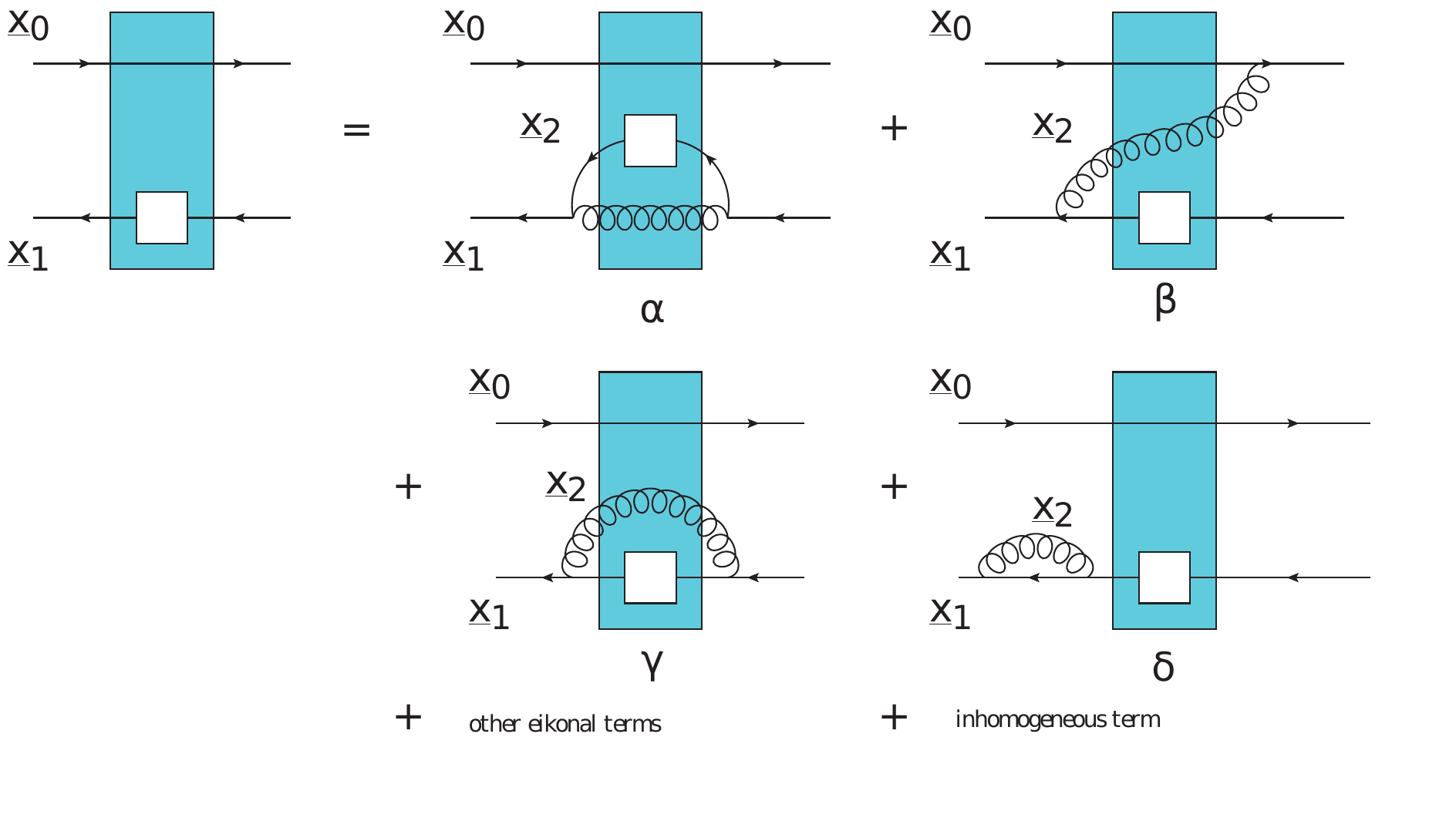}
\caption{The classes of diagrams contributing to the evolution of the amplitudes $G_{10}^{[2]} (z)$ from \eq{seg1t_dips_fin[2]}. The diagram $\alpha$ on the right contains a sub-sub-eikonal soft-quark emission, while the remaining diagrams $\beta, \gamma, \delta$, etc. are a sample of the eikonal emission diagrams \cite{Mueller:1994rr,Mueller:1994jq,Mueller:1995gb,Balitsky:1995ub,Balitsky:1998ya,Kovchegov:1999yj,Kovchegov:1999ua}. }
\label{FIG:H_evolution}
\end{figure}

Having obtained the relevant polarized dipole amplitude for the small-$x$ asymptotics of $g_{1T}^{\textrm{NS}}$, we now turn to its evolution. The only emissions which we need to consider for linearized, large-$N_c$, DLA evolution are unpolarized soft gluon (eikonal) emissions and polarized quark emissions, as shown in \fig{FIG:H_evolution}. From \cite{Kovchegov:2022kyy} we know that the contributions from diagrams of class $\alpha$ vanish due to the tensor structure (the $\un{x}_{10} \cross \un{S}_P$ factor) of the impact parameter integrated polarized dipole amplitude. Thus, we only have the eikonal, unpolarized soft gluon emissions to contend with. These emissions do not carry polarization information, and thus do not allow mixing between different polarized dipole amplitudes. We can see this diagrammatically by looking at the diagrams in classes $\beta$, $\gamma$, and $\delta$ in \fig{FIG:H_evolution}. In each diagram, the white box which symbolizes the polarization dependent sub-eikonal interaction is on the same anti-quark line as it is in the diagram on the LHS of the figure, so the polarized Wilson line entering the dipole amplitude is the same on both sides.

Since there is no mixing of polarized dipole amplitudes through these emissions, a nonzero initial condition for $G_{\un{w} \un{\zeta}}^{[2]}$ cannot generate a nonzero value for any of the other dipole amplitudes. Therefore, $G_{\un{w} \un{\zeta}}^{[2]}$ is the only contributing dipole amplitude for the linearized small-$x$ asymptotics. In the strict DLA, where one only allows for the polarization dependent quark exchange without any eikonal gluons exchanged in the initial condition of the polarized dipole amplitude, it can be shown that the real and virtual eikonal gluon emissions cancel for the leading-$N_c$ initial conditions and evolution \cite{Kovchegov:2016zex}. This would give a trivial evolution equation of the form
\begin{align}
    G_{\un{w} \un{\zeta}}^{[2]} (zs) = G_{\un{w} \un{\zeta}}^{[2] \, (0)} (zs),
\end{align}
with the dipole amplitude being given exactly by its initial condition $G_{\un{w} \un{\zeta}}^{[2] \, (0)}$. Such a cancellation of the class $\alpha$ diagrams also occurred for the sub-sub-eikonal contribution to the Boer-Mulders function \cite{Kovchegov:2022kyy}. In that case the fact that the TMD was T-odd required a higher order initial condition, including an extra gluon exchange beyond the strict DLA. With this extra exchange, the sum over real and virtual eikonal gluon emissions is nonzero and allows for non-trivial small-$x$ evolution. Here we can relax the DLA in the same way for the worm-gear TMD, summing over the eikonal gluon emissions to construct the evolution of the dipole amplitude  $G_{\un{w} \un{\zeta}}^{[2]}$. The resulting equation is identical to those obtained in section III of \cite{Kovchegov:2022kyy}, and yields the linearized, large-$N_c$ asymptotics as
\begin{align}\label{g2_asympts}
    G^{[2]} (x_{10}^2, \zeta \ll 1) \sim \frac{1}{\zeta} J_1 (2 \zeta),
\end{align}
where we have introduced the dimensionless variable $\zeta = \sqrt{\frac{\alpha_s N_c}{2 \pi}} \textrm{ln}(zsx_{10}^2)$. This is an oscillating solution with a decreasing amplitude, so the integral over $z$ in the worm-gear TMD definition \eq{seg1t_fin_simp} will be dominated by the lower limit of the integral and the oscillations will have a negligible effect on the TMD. The resulting asymptotic scaling in $x$ is
\begin{align}
   g_{1T}^{\textrm{NS}} (x \ll 1, k_T^2) \sim x^0.
\end{align}
We find that the flavor non-singlet worm-gear TMD $g_{1T}^{\textrm{NS}}$ at small-$x$ maintains its naive sub-eikonal scaling at DLA, having no dependence on the center of mass energy.


\section{Worm-Gear $h_{1L}^{\perp}$}
\label{sec_h1l}

Now we turn to the worm-gear $h_{1L}^{\perp}$, which can be interpreted as the number density of transversely polarized quarks within a longitudinally polarized proton. It's operator definition is \cite{Meissner:2007rx}
\begin{align} \label{sseh1l_1}
    \frac{k_T^x}{M_P} h_{1L}^{\perp} (x, k_T^2) &= \frac{1}{2} \sum_{S_P} S_P \int \frac{\dd{r}^- \dd[2]{r}_{\perp}}{2(2\pi)^3} \bra{P, S_P} \bpsi (0) \mathcal{U} [0, r] \frac{i \sigma^{1+}\gamma_5}{2} \psi(r)  \ket{P, S_P} \\
    &=\frac{1}{2} \sum_{S_P} S_P \int \frac{\dd{r}^- \dd[2]{r}_{\perp}}{2(2\pi)^3} \bra{P, S_P} \bpsi (0) \mathcal{U} [0, r] \frac{\gamma_5 \gamma^+ \gamma^1}{2} \psi(r)  \ket{P, S_P} \notag
\end{align}
where the proton spin $S_P$ is now in the longitudinal direction. We will again apply the LCOT, leading to an expression analogous to \eq{seg1t_1}, but with $\gamma_5 \gamma^+ \gamma^1$ now projecting out the spinor structure. We need the anti BL spinor product
\begin{align} \label{spin5p1}
    \bar{v}_{\chi_2} (k_2) \frac{\gamma_5 \gamma^+ \gamma^1}{2} v_{\chi_1} (k_1) &= \frac{1}{2 \sqrt{k_1^- k_2^-}} \left[ im \delta_{\chi_1, \chi_2} \un{S} \cross (\un{k}_1 - \un{k}_2) - m \delta_{\chi_1, - \chi_2} \un{S} \vdot (\un{k}_1 + \un{k}_2) \right. \\
    &\left. - \chi_1 \delta_{\chi_1, \chi_2} (2 \un{S} \vdot \un{k}_1 \un{S} \vdot \un{k}_2 - \un{k}_1 \vdot \un{k}_2 - m^2) - i \chi_1 \delta_{\chi_1, - \chi_2} (\un{S} \cross \un{k}_1 \un{S} \vdot \un{k}_2 + \un{S} \cross \un{k}_2 \un{S} \vdot \un{k}_1)  \right] , \notag
\end{align}
where $\un{S} = \hat{x}$ is the direction of spin quantization for the quarks, and the spinors are now in the transverse spin basis which is related to the ordinary anti BL spinors by \cite{Kovchegov:2012ga}
\begin{align}\label{chi_def}
u_\chi \equiv \frac{1}{\sqrt{2}} \, \left[ u_+ + \chi \, u_- \right], \ \ \ v_\chi \equiv \frac{1}{\sqrt{2}} \, \left[ v_+ + \chi \, v_- \right],
\end{align}
where $\chi = \pm 1$. We also need the Wilson line replacement
\begin{align}\label{V_T_repl}
\bar{v}_{\chi_1} (k_1) \left( \hat{V}_{{\un w}}^\dagger \right)^{ji}  v_{\chi_2} (k_2)  \to 2 \sqrt{k_1^- \, k_2^-}  \,   \left( V^{\textrm{pol} \, T \, \dagger}_{{\un{w}}; \chi_2 , \chi_1} \right)^{ji} ,
\end{align}
with
\begin{align}
    V^{\textrm{pol} \, \textrm{T}}_{{\un{x}}; \chi_2 , \chi_1} = \chi \delta_{\chi, \chi'} V_{\un{x}}^{\textrm{T}} + \chi \delta_{\chi, -\chi'} V_{\un{x}}^{\textrm{T} \, \perp} ,
\end{align}
containing polarized Wilson lines defined as
\begin{subequations}\label{VTVT}
\begin{align}
V_{\un{x}}^\textrm{T} \equiv & \, \frac{g^2 \, (p_1^+)^2}{16 \, s^2} \, \int\limits_{-\infty}^{\infty} d{z}_1^- \int\limits_{z_1^-}^\infty d z_2^-  \ V_{\un{x}} [ \infty, z_2^-] \, t^b \, \psi_{\beta} (z_2^-,\un{x}) \, U_{\un{x}}^{ba} [z_2^-,z_1^-] \, \Bigg[ \left[ i \gamma^5 \underline{S} \cdot \cev{\underline{D}}_{x} - \underline{S} \times \cev{\underline{D}}_{x} \right] \, \gamma^+ \gamma^- \\ 
& +  \left[ i \gamma^5 \underline{S} \cdot \underline{D}_{x}  - \underline{S} \times \underline{D}_{x} \right] \gamma^- \gamma^+ \Bigg]_{\alpha \beta} \bar{\psi}_\alpha (z_1^-,\un{x}) \, t^a \, V_{\un{x}} [ z_1^-, -\infty]  , \notag \\
V_{\un{x}}^{\textrm{T} \, \perp} \equiv & \, - \frac{g^2 \, (p_1^+)^2}{16 \, s^2} \, \int\limits_{-\infty}^{\infty} d{z}_1^- \int\limits_{z_1^-}^\infty d z_2^-  \ V_{\un{x}} [ \infty, z_2^-] \, t^b \, \psi_{\beta} (z_2^-,\un{x}) \, U_{\un{x}}^{ba} [z_2^-,z_1^-] \,  \Bigg[ \left[ i \underline{S} \cdot \cev{\underline{D}}_{x} - \gamma^5 \underline{S} \times \cev{\underline{D}}_{x} \right]  \, \gamma^+ \gamma^-  \\ 
& +  \left[ i \underline{S} \cdot \underline{D}_{x}  - \gamma^5 \underline{S} \times \underline{D}_{x} \right] \, \gamma^- \gamma^+ \Bigg]_{\alpha \beta}   \bar{\psi}_\alpha (z_1^-,\un{z}_1) \, t^a \, V_{\un{x}} [ z_1^-, -\infty]  . \notag
\end{align}
\end{subequations}
Taking the limit of massless quarks and performing the same simplifications as in Sec. III of \cite{Kovchegov:2022kyy} yields the leading, sub-sub-eikonal contribution to the worm-gear function as
\begin{align} \label{sseh1l_2}
    \frac{k_T^x}{M_P} h_{1L}^{\perp} (x, k_T^2) &\subset \frac{2 x (p_1^+)^2}{(2\pi)^3} \int \dd[2]{\zeta}_{\perp} \dd[2]{w}_{\perp} \frac{\dd[2]{k}_{1 \perp} \dd{k}_1^- }{(2\pi)^3} \frac{e^{i(\un{k}_1 + \un{k}) \vdot (\un{w} - \un{\zeta})} \theta(k_1^-)}{\un{k}_1^2 \un{k}^2} k_1^- \left( \frac{1}{\un{k}_1^2} + \frac{1}{\un{k}^2} \right) \\
    &\times \left[ (\un{S} \vdot \un{k}_1 \un{S} \vdot \un{k} - \un{k}_1 \vdot \un{k}) \Big{\langle} \tord \tr \left[ V_{\un{\zeta}} V_{\un{w}}^{\textrm{T} \dagger} \right] + \atord \tr \left[ V_{\un{w}}^{\textrm{T}} V_{\un{\zeta}}^{\dagger} \right] \Big{\rangle} \right. \notag \\
    &\left. + i ( \un{S} \cross \un{k}_1 \un{S} \vdot \un{k} + \un{S} \vdot \un{k}_1 \un{S} \cross \un{k}) \Big{\langle} \tord \tr \left[ V_{\zeta} V_{\un{w}}^{\textrm{T} \perp \dagger} \right] - \atord \tr \left[  V_{\un{w}}^{\textrm{T} \perp}  V_{\un{\zeta}}^{\dagger} \right]\Big{\rangle} \right] . \notag
\end{align}
We find the same polarized Wilson line correlators as in the case of the Boer-Mulders TMD, but with different tensor structures multiplying them. Subtracting the anti-quark TMD contribution yields the flavor non-singlet worm-gear TMD as
\begin{align}\label{sseh1l_3}
& \frac{k_T^x}{M_P} h_{1L}^{\perp \ \textrm{NS}} (x, k_T^2) =  \frac{i x N_c}{\pi^3} \int d^2 {x_{10}} \, \frac{ d^2{k_{1 \perp}} }{(2\pi)^3} \, e^{i (\underline{k}_1 + \underline{k} ) \cdot {\un x}_{10}  } \, \frac{1}{\underline{k}_1^2 \, \underline{k}^2} \left(  \frac{1}{\underline{k}_1^2} +  \frac{1}{\underline{k}^2}  \right)   \int\limits_\frac{\Lambda^2}{s}^1 \frac{dz}{z} \\
\times & \Big\{  (2 \, {\un S} \cdot {\un k}_1 \, {\un S} \cdot {\un k} - {\un k}_1 \cdot {\un k} ) \, H^{1}_{10} (z )  + ({\un S} \times {\un k}_1 \, {\un S} \cdot {\un k} + {\un S} \cdot {\un k}_1 \, {\un S} \times {\un k} )  \, H^{2}_{10} (z ) \Big\} , \notag
\end{align}
where we have the polarized dipole amplitudes 
\begin{subequations}\label{H_defs}
\begin{align}
& H^{1}_{10} (z) \equiv \frac{1}{2 N_c} \, \mbox{Im} \,  \llangle \tord \tr \left[ V_{\underline{0}} \, V^{\textrm{T} \, \dagger}_{{\un 1}} \right] - \tord \tr \left[ V_{\underline{0}}^\dagger \, V^{\textrm{T} }_{{\un 1}} \right] \rrangle, \\
& H^{2}_{10} (z) \equiv \frac{1}{2 N_c} \, \mbox{Re} \,  \llangle \tord \tr \left[ V_{\underline{0}} \, V^{\textrm{T} \, \perp \, \dagger}_{{\un 1}} \right] - \tord \tr \left[ V_{\underline{0}}^\dagger \, V^{\textrm{T} \, \perp}_{{\un 1}} \right]  \rrangle , 
\end{align}
\end{subequations}
with the double angle brackets now scale out two factors of energy, $\llangle ... \rrangle = (zs)^2 \Big{\langle} ... \Big{\rangle}$, and where again $\un{x}_{10} = \un{x}_1 - \un{x}_0 = \un{w} - \un{\zeta}$.
Noting  that there is no $\epsilon^{ij}$ tensor structure on the LHS, we can see that the $H^2_{10}$ term on the RHS needs to come in with an extra $\epsilon^{ij}$ after impact parameter integration, while the $H^1_{10}$ term should only pick up a dot product term. We conclude that the impact parameter integrals of these two dipole amplitudes must have the structure
\begin{subequations}\label{HS_tens}
\begin{align}
& \int d^2 b_\perp \, H^{1}_{10} (z) = {\un x}_{10} \times {\un S} \, H^{1} (x_{10}^2, z ) ,  \label{HS_tensa} \\
& \int d^2 b_\perp \, H^{2}_{10} (z) = {\un x}_{10} \cdot {\un S} \, H^{2} (x_{10}^2, z ) , \label{HS_tensb} 
\end{align}
\end{subequations}
exactly as in the Boer-Mulders case \cite{Kovchegov:2022kyy}.

The evolution of these dipole amplitudes is again given the diagrams in \fig{FIG:H_evolution}, and the same arguments as we applied above and as were applied for the Boer-Mulders TMD in \cite{Kovchegov:2022kyy} lead us to consider sub-leading $N_c$ initial conditions which will be driven to small-$x$ by eikonal gluon emissions. The resulting small-$x$ asymptotics are
\begin{align}\label{h_asympts}
    H^1 (x_{10}^2, \zeta \ll 1) = H^2 (x_{10}^2, \zeta \ll 1) \sim \frac{1}{\zeta} J_1 (2 \zeta),
\end{align}
which gives the TMD scaling as
\begin{align}
    h_{1L}^{\perp \textrm{NS}} (x \ll 1, k_T^2) \sim \left( \frac{1}{x} \right)^{-1}.
\end{align}
We find that, as with the Boer-Mulders TMD and the $g_{1T}$ worm-gear TMD above, the flavor non-singlet worm-gear $h_{1L}^{\perp}$ TMD at small-$x$ maintains its naive sub-sub-eikonal scaling at DLA, falling off linearly with $x$.


\section{Pretzelosity $h_{1T}^{\perp}$}
\label{sec_pretz}

Finally we turn to the pretzelosity distribution, which has an operator definition as
\begin{align} \label{pretz_def1}
    \frac{k_T \vdot S_P \, k_T \cross S_P}{M_P^2} h_{1T}^{\perp \, q} (x, k_T^2) = \epsilon^{ij} S_P^i  \int \frac{\dd{r}^- \dd[2]{r}_{\perp}}{2(2 \pi)^3} \bra{P,S_P} \bar{\psi} (0) \mathcal{U} [0, r] \frac{i \sigma^{j+} \gamma_5}{2} \psi (r) \ket{P, S_P},
\end{align}
where we can explicitly take the proton spin along the $y$-direction, $S_P = \hat{y}$, to simplify to
\begin{align} \label{pretz_def2}
    -\frac{k_T^x k_T^y}{M_P^2} h_{1T}^{\perp \, q} (x, k_T^2) = \int \frac{\dd{r}^- \dd[2]{r}_{\perp}}{2(2 \pi)^3} \bra{P,S_P} \bar{\psi} (0) \mathcal{U} [0, r] \frac{\gamma_5 \gamma^+ \gamma^1 }{2} \psi (r) \ket{P, S_P}.
\end{align}
Once again the Dirac structure is projected out by $\gamma_5 \gamma^+ \gamma^1$, so we have the leading small-$x$ contribution coming from the same sub-sub-eikonal structure as above in the worm-gear $h_{1L}^{\perp}$ (\eq{sseh1l_2}). Going straight to the flavor non-singlet TMD, we have
\begin{align}\label{pretz_def3}
&-\frac{k_T^x k_T^y}{M_P^2} h_{1T}^{\perp \, q \, \textrm{NS}} (x, k_T^2) \Big{|}_{\textrm{sub-sub-eik}}  \subset  \frac{i x N_c}{\pi^3} \int d^2 {x_{10}} \, \frac{ d^2{k_{1 \perp}} }{(2\pi)^3} \, e^{i (\underline{k}_1 + \underline{k} ) \cdot {\un x}_{10}  } \, \frac{1}{\underline{k}_1^2 \, \underline{k}^2} \left(  \frac{1}{\underline{k}_1^2} +  \frac{1}{\underline{k}^2}  \right)   \int\limits_\frac{\Lambda^2}{s}^1 \frac{dz}{z} \\
&\times \Big\{  (2 \, {\un S} \cdot {\un k}_1 \, {\un S} \cdot {\un k} - {\un k}_1 \cdot {\un k} ) \, H^{1 T} ( x_{10}, z )  + ({\un S} \times {\un k}_1 \, {\un S} \cdot {\un k} + {\un S} \cdot {\un k}_1 \, {\un S} \times {\un k} ) \, H^{2 T} ( x_{10}, z ) \Big\} . \notag
\end{align}
The dipole amplitudes entering here are defined again as
\begin{subequations}\label{pretz_dipoles}
\begin{align}
& H^{1 T} (x_{10}, z) \equiv \frac{1}{2 N_c} \, \mbox{Im} \,  \llangle \tord \tr \left[ V_{\underline{0}} \, V^{\textrm{T} \, \dagger}_{{\un 1}} \right] - \tord \tr \left[ V_{\underline{0}}^\dagger \, V^{\textrm{T} }_{{\un 1}} \right] \rrangle, \\
& H^{2 T} (x_{10}, z) \equiv \frac{1}{2 N_c} \, \mbox{Re} \,  \llangle \tord \tr \left[ V_{\underline{0}} \, V^{\textrm{T} \, \perp \, \dagger}_{{\un 1}} \right] - \tord \tr \left[ V_{\underline{0}}^\dagger \, V^{\textrm{T} \, \perp}_{{\un 1}} \right]  \rrangle , 
\end{align}
\end{subequations}
with double angle brackets scaling out $(zs)^2$. The LHS of \eq{pretz_def3} is even under $k_T \rightarrow -k_T$, so integrating the dipole amplitudes over impact parameter cannot generate terms linear in $\un{x}_{10}$ coupling to the spin vectors. This means that the diagrams in class $\alpha$ from \fig{FIG:H_evolution} do not vanish. In the strict DLA limit the eikonal diagrams vanish, and we have the same contribution to evolution as for the polarized dipole amplitude contributing to the transversity TMD in \cite{Kovchegov:2018zeq}, with the class $\alpha$ diagrams yielding the linearized, large-$N_c$, DLA evolution for the dipole amplitudes as
\begin{subequations}
    \begin{align}
        H^{1 T}_{10} (zs) = H^{1 T \, (0)}_{10} (zs) + \frac{\as N_c}{2\pi} \int\limits_{\Lambda^2/s}^z \frac{\dd{z}'}{z'} \int\limits_{1/z's}^{x_{10}^2 z/z'} \frac{\dd{x}_{21}^2}{x_{21}^2} H^{1 T}_{21} (z's), \\
        H^{2 T}_{10} (zs) = H^{2 T \, (0)}_{10} (zs) + \frac{\as N_c}{2\pi} \int\limits_{\Lambda^2/s}^z \frac{\dd{z}'}{z'} \int\limits_{1/z's}^{x_{10}^2 z/z'} \frac{\dd{x}_{21}^2}{x_{21}^2} H^{2 T}_{21} (z's),
    \end{align}
\end{subequations}
where we have introduced the impact parameter integrated dipole amplitudes as 
\begin{subequations} \label{pretz_dips_def}
    \begin{align}
        H^{1 T}_{10} (zs) = \int \dd[2]{b}_{\perp} H^{1 T} (x_{10}, zs) , \\
        H^{2 T}_{10} (zs) = \int \dd[2]{b}_{\perp} H^{2 T} (x_{10}, zs) .
    \end{align}
\end{subequations}
This equation can be solved analytically (cf. \cite{Kovchegov:2018zeq}), yielding the asymptotics for the flavor non-singlet pretzelosity TMD as
\begin{align} \label{pretz_fin}
    h_{1 T}^{\perp \textrm{NS}} (x \ll 1, k_T^2) \sim \left( \frac{1}{x} \right)^{-1 + 2 \sqrt{\frac{\as N_c}{2\pi}}},
\end{align}
exactly matching the scaling of the corresponding transversity TMD.


\section{Conclusions}
\label{sec_conc}

In this paper we studied the small-$x$ asymptotics of the flavor non-singlet, leading-twist spin-spin coupling TMDs: the two worm-gear functions $g_{1T}$ and $h_{1L}^{\perp}$, and the pretzelosity $h_{1T}^{\perp}$. We applied the LCOT, rewriting the operator definition of each TMD in the small-$x$ limit in terms of polarized dipole amplitudes, and found that all three TMDs reduced essentially to polarized dipole amplitudes which are known from previous studies of other leading-twist TMDs. The close similarity of the contributing polarized dipole amplitudes means that the DLA, large-$N_c$, linearized small-$x$ evolution for all three TMDs are known equations with known solutions, either through exact solutions or numerical calculations. We began with the flavor non-singlet worm-gear TMD $g_{1T}^{\textrm{NS}}$, showing that it reduces to almost the same dipole amplitudes as the sub-eikonal contribution to the quark Sivers function \cite{Kovchegov:2022kyy}, given in \eq{seg1t_dips}. The differences between the dipole amplitudes contributing to this worm-gear TMD and the Sivers function yield a substantial change by forcing most of the dipole amplitudes to have zero initial conditions, and the small-$x$ asymptotics are controlled by polarized quark exchange operators. The tensor structure of the impact parameter integrated dipole amplitude makes the strict DLA small-$x$ evolution trivial. Thus, we were led to consider the same eikonal gluon emission driven evolution as studied for the Boer-Mulders function \cite{Kovchegov:2022kyy}, where going beyond the strict DLA limit was required by the T-odd nature of the TMD. This evolution yields an oscillating solution, where the oscillations of the dipole ultimately are washed out upon integrating over the internal momentum fraction variable $z$ as in \eq{seg1t_fin_simp}. Thus, the scaling of the TMD is unchanged from its naive sub-eikonal scaling, and we find
\begin{align}\label{asympts_g1t}
    g_{1T}^{\textrm{NS}} (x \ll 1, k_T^2) = C_T(x, k_T^2) \left( \frac{1}{x} \right)^0 + ... ,
\end{align}
where the ellipses represent corrections which would come in at sub-sub-eikonal order. 

Next we considered the flavor non-singlet worm-gear TMD $h_{1L}^{\perp \, \textrm{NS}}$, which we found reduces to sub-sub-eikonal polarized dipole amplitudes \eq{H_defs} analogous to those which appear in the Boer-Mulders function. As with $g_{1T}^{\textrm{NS}}$, the tensor structure of the impact integrated dipole amplitudes (as shown in \eq{HS_tens}) suggested that we consider the eikonal gluon emission driven evolution as studied for the Boer-Mulders function \cite{Kovchegov:2022kyy}, yielding the same oscillating with decreasing amplitude behavior. Thus, the scaling of the TMD is unchanged from its naive sub-sub-eikonal scaling, and we find
\begin{align}
    h_{1L}^{\perp \textrm{NS}} (x \ll 1, k_T^2) = C_L(x, k_T^2) \left( \frac{1}{x} \right)^{-1} + ...
\end{align}
It is important to note that there is a potential finite light quark mass correction proportional to $m / k_T$ entering in at sub-eikonal order. The dipole amplitudes entering are again the same as in the case of the Boer-Mulders function, so the contributions would likely not evolve and yield an additive correction from the initial conditions. We leave a detailed investigation of the finite quark mass corrections for future work. 
The scaling of the evolved worm-gear TMDs $g_{1T}^{\textrm{NS}}$ and $h_{1L}^{\perp \, \textrm{NS}}$ as $x^0$ and $x$ respectively even after evolution is a very interesting feature. Based on the spin-dependent odderon contribution to the eikonal quark Sivers function \cite{Szymanowski:2016mbq,Dong:2018wsp,Kovchegov:2021iyc,Kovchegov:2022kyy} and several gluon TMDs \cite{Boer:2015pni}, which is known to maintain eikonal $(1/x)$ scaling under the effects of linear evolution \cite{Bartels:1999yt,Kovchegov:2003dm,Kovchegov:2012rz,Caron-Huot:2013fea,Brower:2008cy,Avsar:2009hc,Brower:2014wha}, as well as the evolution of the sub-sub-eikonal Boer-Mulders TMD which maintains linear $x$ scaling under linearized evolution, one might conjecture that there is some protection against evolution corrections coming from the T-odd property of all of these TMDs. With the two worm-gear functions we have examples of where the evolution effects again do not alter the naive sub-(sub-)eikonal scaling as $x$ for T-even TMDs, so if there is an underlying symmetry protecting these TMDs from evolution it is likely not determined by T-parity alone. 

Finally, we studied the flavor non-singlet pretzelosity TMD, where the small-$x$ asymptotics came from polarized dipole amplitudes (\eq{pretz_dipoles}) analogous to the one entering the evolution of the quark transversity TMD. The evolution equations once again are already known, yielding the DLA, large-$N_c$, linearized asymptotics as 
\begin{align}
    h_{1T}^{\perp \textrm{NS}} (x \ll 1, k_T^2) = C_{T \perp} (x, k_T^2) \left( \frac{1}{x} \right)^{-1 + 2 \sqrt{\frac{\as N_c}{2\pi}}} + ...
\end{align}
This is the same scaling as found in \cite{Kovchegov:2018zeq} for the flavor non-singlet quark transversity TMD, which is a linear combination of $h_{1T}^{\textrm{NS}}$ and $h_{1T}^{\perp \textrm{NS}}$, so indeed one would anticipate a match between the dipole amplitudes and consequently the asymptotics. Again, there is a potential finite light quark mass correction to these results.

Combining the results in this study with those obtained for the quark helicity TMD \cite{Kovchegov:2015pbl,Kovchegov:2017lsr, Kovchegov:2018znm, Cougoulic:2022gbk,Borden:2023ugd}, the quark transversity TMD \cite{Kovchegov:2018zeq}, the quark Sivers and Boer-Mulders TMDs \cite{Boer:2015pni,Szymanowski:2016mbq,Dong:2018wsp,Kovchegov:2021iyc,Kovchegov:2022kyy}, as well as the asymptotics describing the unpolarized quark TMD \cite{Itakura:2003jp,Ermolaev:1995fx}, we have known small-$x$ asymptotics for each of the leading twist quark TMDs, collected in \tab{tab_tmds}. Let's note a few interesting features of the table: firstly, the diagonal entries all have small-$x$ scaling nearly equal to that of the Reggeon up to a factor of $x$. There appears to be some universality among the TMDs which survive $k_T$ integration to yield the three flavor non-singlet parton distribution functions (PDFs), although the pretzelosity TMD does not precisely fit in with this characterization. For the off diagonal entries, the leading terms are all exactly their naive scaling based on the quark spin content. The Sivers function is an unpolarized quark density and thus is eikonal, while the worm-gear $g_{1T}$ is sub-eikonal due to probing the quarks' helicity, and the Boer-Mulders TMD and the worm-gear $h_{1L}^{\perp}$ are both sub-sub-eikonal due to probing the quarks' transversity. It may be that there is a symmetry protecting these off-diagonal TMDs from evolution effects based on the coupling between different directions of spin, as the quarks in these TMDs are all polarized in a different direction than their parent hadron (for example quark helicity with hadron transversity for $g_{1T}$).

Several of the leading-twist quark TMDs only have their small-$x$ asymptotics calculated for the flavor non-singlet functions, the flavor singlet functions which can mix with the gluon TMDs under evolution are a topic for future work. The flavor non-singlet asymptotics provide constraints on the valence quark distributions, giving crucial input for future TMD studies at the EIC \cite{Accardi:2012qut,Boer:2011fh,Proceedings:2020eah,AbdulKhalek:2021gbh}. Furthermore, the operator equations for the polarized dipole amplitudes have been constructed in the large-$N_c$ limit, so one could go beyond the linearized DLA approximation used here to obtain precise phenomenological results. Going beyond the large-$N_c$ limit, one could consider the large-$N_c$ and $N_f$ limit to restore quark contributions \cite{Kovchegov:2018znm,Kovchegov:2020hgb,Cougoulic:2022gbk,Adamiak:2023okq} where they have been neglected (for example in the sub-eikonal contribution to the Sivers function \cite{Kovchegov:2022kyy}). Obtaining evolution equations for all $N_c$ values requires a generalization of the Jalilian-Marian-Iancu-McLerran-Weigert-Leodinov-Kovner (JIMWLK) equations \cite{Jalilian-Marian:1997jx,Jalilian-Marian:1997gr,Jalilian-Marian:1997dw,Iancu:2001ad,Iancu:2000hn} to include the sub-eikonal and sub-sub-eikonal operators needed for quark helicity dependent and quark transversity dependent insertions within the Wilson lines, as has been studied in \cite{Cougoulic:2019aja,Cougoulic:2020tbc}.


\section*{Acknowledgements}

The author is thankful to Yuri Kovchegov for enlightening discussions on this work and helpful comments on the manuscript. This work was supported by the Center for Nuclear Femtography, Southeastern Universities Research Association, Washington, D.C..


%

\end{document}